\documentclass[reprint,amsmath,amssymb,notitlepage]{revtex4-1}
\pdfoutput=1

\usepackage{graphicx}
\usepackage{hyperref}

\makeatletter
\hypersetup{
	    pdfcreator={Cesar Henrique Comin},
		colorlinks=true,       		
    	linkcolor=black,          	
    	citecolor=black,        		
    	filecolor=black,      		
		urlcolor=black,
		bookmarksdepth=4
}
\makeatother

\begin{document}

\title{A diffusion-based approach to obtaining the borders of urban areas}

\author{Cesar Henrique Comin$^1$}
\email{Email: chcomin@gmail.com}
\author{Filipi Nascimento Silva$^1$}
\author{Luciano da Fontoura Costa$^1$}

\affiliation{$^1$Institute of Physics at S\~ao Carlos, University of S\~ao Paulo, S\~ao Carlos, S\~ao Paulo, Brazil}

\begin{abstract}
The access to an ever increasing amount of information in the modern world gave rise to the development of many quantitative indicators about urban regions in the globe. Therefore, there is a growing need for a precise definition of how to delimit urban regions, so as to allow proper respective characterization and modeling. Here we present a straightforward methodology to automatically detect urban region borders around a single seed point. The method is based on a diffusion process having street crossings and terminations as source points. We exemplify the potential of the methodology by characterizing the geometry and topology of 21 urban regions obtained from 8 distinct countries. The geometry is studied by employing the lacunarity measurement, which is associated to the regularity of holes contained in a pattern. The topology is analyzed by associating the betweenness centrality of the streets with their respective class, such as motorway or residential, obtained from a database.
\end{abstract}

\maketitle

\section{Introduction}

The characterization and modeling of urban areas~\footnote{We note that, in this study, the terms \emph{urban area} and \emph{metropolitan area} are considered interchangeable. They may have strong distinctions for some countries.} stands as an important aspect of improving, in an efficient manner, their infrastructure, transportation and growth planning. Among many distinct properties of urban areas that can be studied, universal scaling laws of urban area sizes and population have received particular interest in the literature~\cite{gabaix2004evolution,saichev2009theory}. Special attention is given to the long standing problem of the validity and explanation of Zipf's law~\cite{zipf1949human,gabaix1999zipf,jiang2011zipf}, and its conciliation with Gibrat's law of proportionate growth~\cite{eeckhout2004gibrat,levy2009gibrat,gabaix1999zipf}. Strikingly, most studies about urban areas are not concerned with the proper definition of their borders, a problem which is related to the very definition of an urban area.

\begin{figure}[]
  \begin{center}
 \includegraphics[width=0.95\columnwidth]{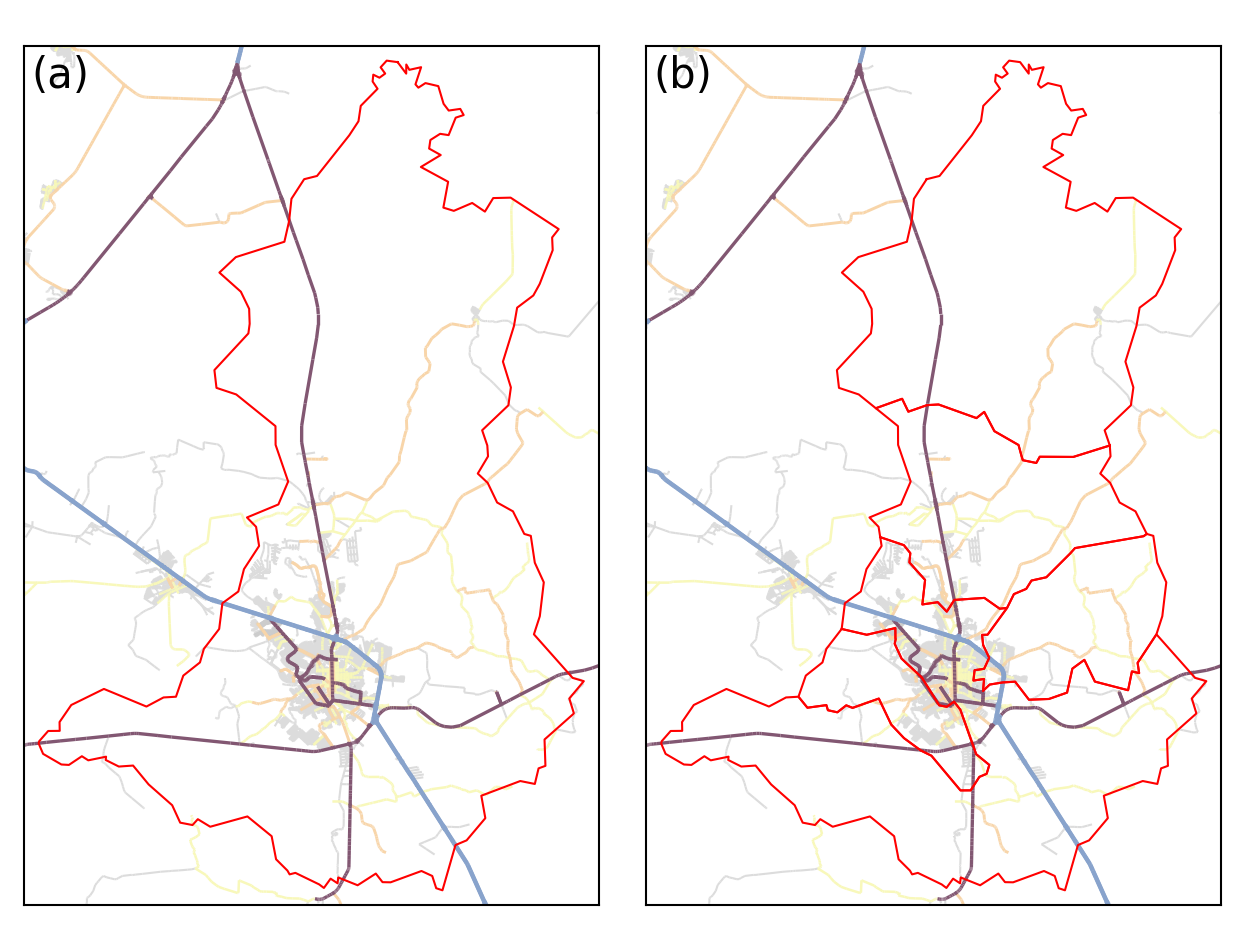}
  \caption{Example of administrative regions visualized on the map of a city. The administrative regions at the (a) city and (b) district levels are shown in red. The main inhabited region of the city can be seen at the lower part of the map.}
  ~\label{f:adm_region}
  \end{center}
\end{figure}

It is indeed a difficult task to provide a general definition of urban area borders~\cite{gabaix1999zipf,gabaix2004evolution,krugman1996self}. A common approach in network theory is to use the main administrative region of the cities constituting the urban area of interest, which are compiled in great detail in the Global Administrative Areas dataset~\footnote{ Available at http://www.gadm.org/}. However, administrative regions have a number of problems. First, in many cases their delimitation does not have a clear motivation~\cite{auffhammer2009exploring,alesina2000political}. Second, administrative areas can have different meanings for distinct countries, in the sense that the territories may be divided at different scales (e.g. province, state, county or municipality) for administration purposes. Third, in many cases they are unrelated to the population density, that is, cities with low population tend to have administrative regions composed by a small inhabited center and a large, mostly uninhabited, area. Such uninhabited areas commonly contain a sparse distribution of infrastructure, which is a potential source of bias when characterizing the regions. In Figure \ref{f:adm_region} we show an example of such a problem. The official administrative region of the city, shown in Figure \ref{f:adm_region}(a), is much larger than the actual inhabited region of the city. Using official subdivisions of such an administrative area (shown in Figure \ref{f:adm_region}(b)) is also a source of problems, since the city can become divided into regions that should be considered as one.

Other definitions can be used to delimit urban areas~\cite{holmes2010cities,rozenfeld2008laws}. An overview of the methodologies adopted by a number of countries can be found in~\cite{uniteddemographic2005}. Unfortunately, such methodologies are usually aimed at capturing specific characteristics observed in cities belonging to countries where the criteria was developed. Besides, they usually rely on manual interventions to set the main regions of interest or to avoid outlier cases. Furthermore, since the majority of the methods are based on population density, they are influenced by the census methodology adopted at each region. For example, a city might be divided into large districts for census data organization purposes, and a single value of population density be associated to each district. The respective border found using such data will be influenced by the district borders.

In this work we present a general procedure that can be applied to any urban region, regardless of the actual definition of border used for each specific country. The procedure is based on the structure of the urban area streets, which composes the majority of the transportation network of such systems. The use of an urban area street configuration has a clear motivation, since the density of streets in an urban area is known to be correlated with many other indicators, the most often considered ones being urbanization and population density~\cite{makse1995modelling,rozenfeld2009area,murcio2013second}. Some methodologies to define borders using streets have been described in the literature~\cite{jiang2011zipf,masucci2015problem,jiang2015zipf}. Our approach differs from the other studies by considering a distinct type of clustering process for the street crossings. Also, the three parameters employed in the methodology can be adjusted to follow distinct qualitative and quantitative rules setting the criteria for urban area borders.

Two applications of the methodology are also provided. We obtained 21 urban regions from 8 countries and analyzed their street patterns geometrically and topologically. The geometry of the urban areas is analyzed through the lacunarity measurement~\cite{gefen1983geometric,allain1991characterizing}, a traditional approach to characterize fractal structures. The edge betweenness centrality~\cite{brandes2008variants} is used to infer the expected traffic flow in the topology of each street network.

\section{Defining the borders of urban areas}

\begin{figure}[]
  \begin{center}
 \includegraphics[width=0.95\columnwidth]{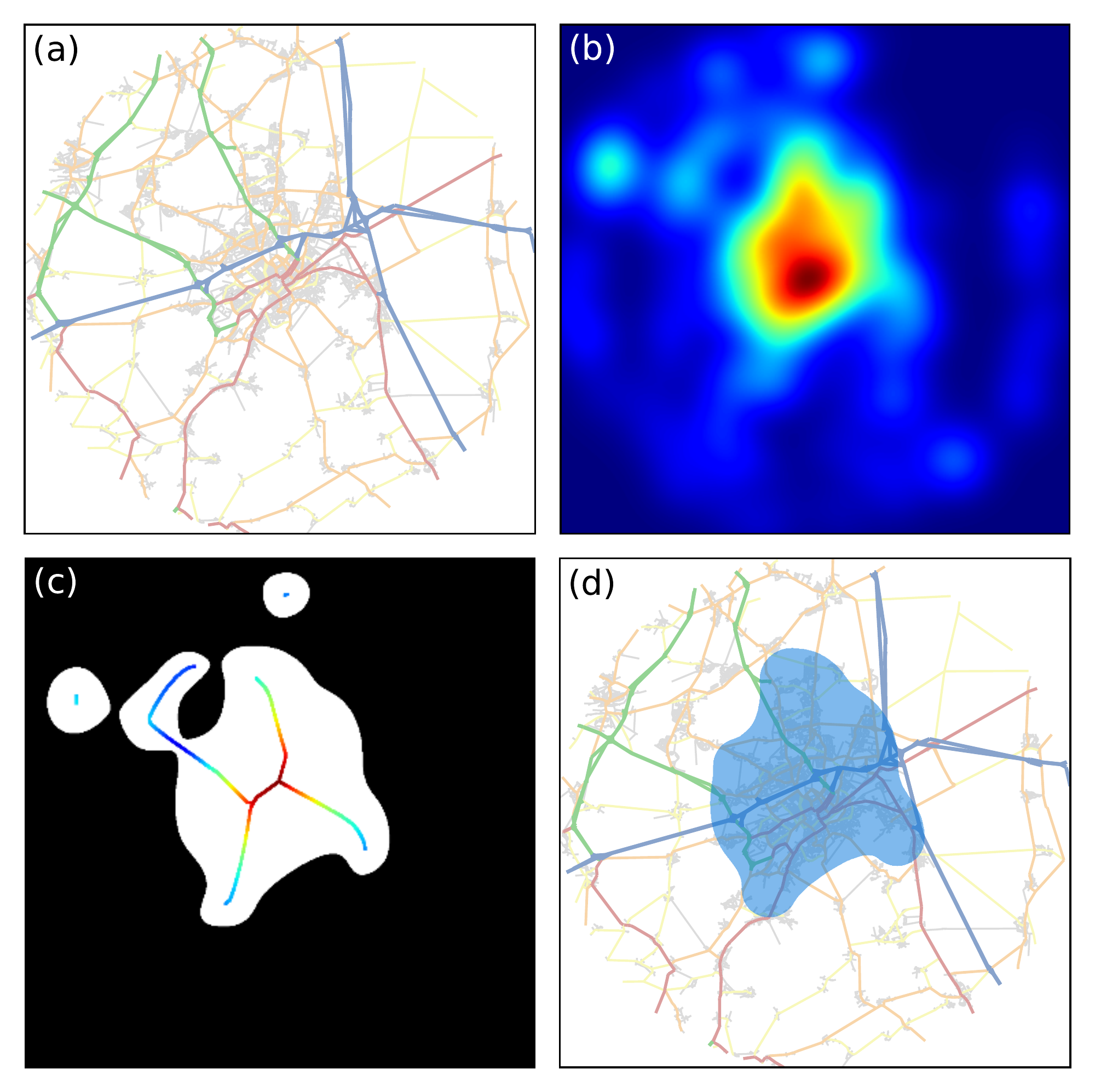} 
  \caption{Steps of the proposed methodology to identify urban area borders. (a) Streets retrieved in a circle of radius $R$ around a seed point. (b) Probability density distribution of street crossings and terminations. (c) Candidate regions obtained after applying a threshold to the probability density. The skeleton of the region is also shown. Each skeleton point contains the distance to the closest border point. (d) Urban region obtained after applying a threshold to the skeleton, and retrieving the region with the largest area.}
  ~\label{f:city_region}
  \end{center}
\end{figure}

The first step of the method is to define a reference point inside the urban area, for geolocation purposes. It is best, but not strictly necessary, that this point be close to the center of the main city of the urban area. The city centers required for this step can be obtained from the Wikipedia~\footnote{https://www.wikipedia.org/} page for each urban region being analyzed. Alternatively, one can obtain the centers from the GeoNames dataset~\footnote{Available at http://www.geonames.org/}. All street intersections or endings that are in a radius $R$ from the reference point are retrieved. In Figure \ref{f:city_region}(a) we show an example of the initial streets one might obtain in this step. The urban area is then described by a graph, where street intersections and terminations are represented by nodes and two nodes are connected by an edge whenever there is a street between them. Next, a probability density, $P$, is estimated by propagating street intersections and endings inside the region. This propagation is done by considering the intersections and endings as sources of a diffusion process taking place inside the initial region. This diffusion scheme allows to estimate what we call the structural density of the city. Therefore, larger probabilities are associated with city regions having more street terminations and intersections.  Computationally, this physical process is equivalent to applying a kernel density estimation~\cite{wand1994kernel} to the initial set of street intersections and terminations. For such a task, we use a Gaussian kernel with standard deviation $\sigma$. The density map for the streets displayed in Figure \ref{f:city_region}(a) is shown in Figure \ref{f:city_region}(b). We note that parameter $\sigma$ is related to the time allowed for the diffusion to take place. This parameter sets the scale at which we consider that nearby streets belong to the same urban area.

Candidate urban regions are then defined by applying a threshold, $T$, to the structural density and selecting regions where the density is larger than $T$. The threshold is defined as a fraction of the maximum value of the structural density, that is, $T=fP_{\mathrm{max}}$. Then, the skeletons~\cite{da2009shape}, or medial axis, of the candidate regions are calculated. The shortest distance between each skeleton point, $s$, and the candidate region border is stored as $D(s)$. In Figure \ref{f:city_region}(c) we show in white the regions found after applying the threshold to the density map, together with the distances $D(s)$ associated to each skeleton point. The skeleton is used as a means to separate nearby cities from the main urban area. If a city is close to the region we are interested in, but there are few streets between the city and the region, $D(s)$ will have a small value. On the other hand, if the nearby city is so close to the region of interest that their streets do not present a clear separation, we consider that its street network belongs to the main urban area. Therefore, a skeleton pruning procedure is done by applying a morphological dilation~\cite{da2009shape} to the obtained skeletons using a disk with radius $D(s)$ as a structuring element, but only skeleton points $s$ where $D(s)>M_w$ are dilated. The parameter $M_w$ sets the minimum width of the desired urban region at the interface between different cities. More than one region might be created by the previous step, the region with the largest physical area becomes the relevant urban area. The resulting region for our illustrative example is shown in Figure \ref{f:city_region}(d). We note that a hard boundary can be added to the methodology in order to separate urban regions belonging to distinct countries.

\begin{figure}[]
  \begin{center}
 \includegraphics[width=0.95\columnwidth]{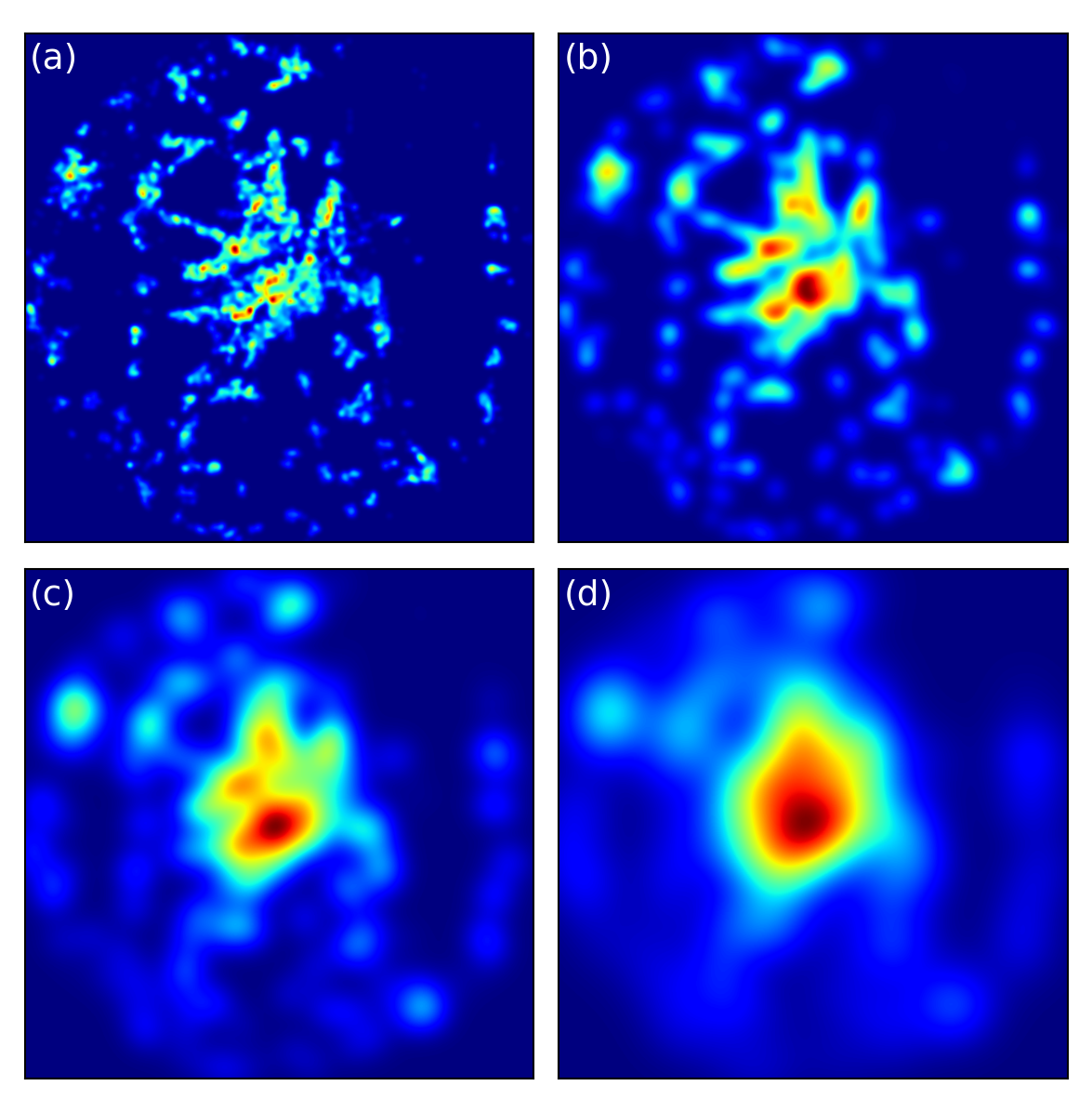} 
  \caption{The influence of the diffusion time scale $\sigma$ on the predicted structural density of a city. Each panel shows the resulting density when setting (a) $\sigma=0.19$ km, (b) $\sigma=0.56$ km, (c) $\sigma=1.11$ km, (d) $\sigma=1.86$ km.}
  ~\label{f:mulhouse_diffusion}
  \end{center}
\end{figure}

The parameters $\sigma$, $f$ and $M_w$ should be the same for all considered urban areas. This is because their value provide the actual definition of an urban area. That is, we consider that urban areas should be formed by streets that are, roughly, closer than $\sigma$, have a density of intersections larger than $f$ and do not form patches thinner than $M_w$. The main parameter of the method is the diffusion time $\sigma$, which sets the scale of the structural density. In Figure \ref{f:mulhouse_diffusion} we show  the structural density of the candidate region displayed in Figure \ref{f:city_region}(a) for distinct values of $\sigma$. When $\sigma$ is small, the diffusion dynamics only reaches the immediate neighborhood of street intersections and terminations, as shown in Figure \ref{f:mulhouse_diffusion}(a). Such scale can be useful for city planning purposes, as the structural density can be related to the expected traffic flow at each respective region, due to its relationship with the density of street crossings. At intermediate scales such as those shown in Figures \ref{f:mulhouse_diffusion}(b) and (c), the structural density can be used to define subdivisions of a city for administration or characterization purposes. Nevertheless, our interest lies on a sufficiently long time of the diffusion process, where the influence of structural variations inside the city becomes negligible, as shown in Figure \ref{f:mulhouse_diffusion}(d).

\begin{figure*}[]
  \begin{center}
 \includegraphics[width=\linewidth]{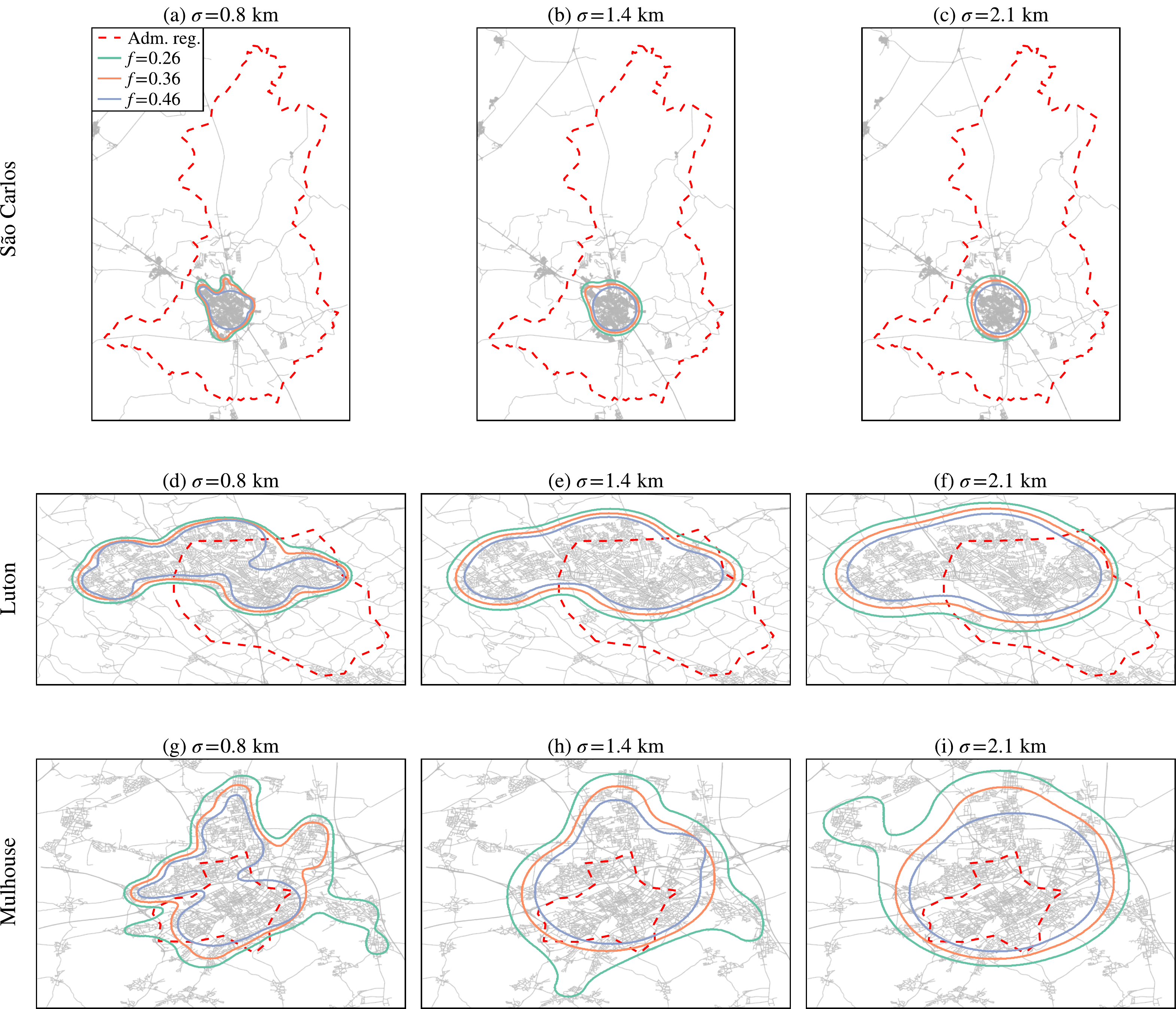} 
  \caption{Borders identified by the proposed methodology, compared with administrative regions defined for three cities. Two parameters of the methodology were varied. The values used for parameter $f$ are indicated in panel (a). Parameter $\sigma$ was set to the values indicated above each figure. The considered cities are S\~ao Carlos (Figures (a), (b) and (c)), Luton (Figures (d), (e) and (f)) and Mulhouse (Figures (g), (h) and (i)).}
  ~\label{f:param_variation}
  \end{center}
\end{figure*}

In order to illustrate the potential of using the structural density to establish urban area borders, we compare the borders defined according to administrative regions with those obtained from the presented methodology. In Figure~\ref{f:param_variation} we show maps of the street networks of three cities, and the respective administrative regions. We also show the urban borders obtained by setting different values for the parameters of the methodology. The distinct borders shown in each map correspond to different respective values of the parameter $f$, which are indicated in the legend of Figure~\ref{f:param_variation}(a). Parameter $\sigma$ was also varied, the considered values being indicated above each figure. Parameter $M_w$ was kept fixed at $M_w=100$ m for all cases. 

Regarding the city of S\~ao Carlos, since its streets are concentrated in a specific region, with few streets outside it, the parameters of the method have little impact on the obtained border, as can be seen in Figures~\ref{f:param_variation}(a), (b) and (c). It is also clear that the obtained borders are more intuitive than the administrative region defined for the city, which is much larger than the urban part of this city. For the city of Luton, we observe in Figure~\ref{f:param_variation}(d) that setting a small $\sigma$ and large $f$ provides a border that seems too small to accommodate the city's street network, while larger values of these parameters gives a more intuitive border for the urban area of the city, as shown in Figures~\ref{f:param_variation}(e) and (f). A particularly interesting result is obtained for the city of Mulhouse. The borders shown in Figure~\ref{f:param_variation}(g) indicate that, by setting a sufficiently small value of $\sigma$, parameter $f$ can be changed to define different core regions of the street network of the city. That is, large values of $f$ define core regions having a disproportionately high density of streets. This situation is not observed when setting larger values of $\sigma$, as shown in Figures~\ref{f:param_variation}(h) and (i).


In the following, we provide examples of studies that can be applied to urban regions identified by the methodology. We consider 21 urban regions belonging to 8 different countries, obtained using the OpenStreetMap Overpass API~\footnote{http://wiki.openstreetmap.org/wiki/Overpass\_API}. The name and location of each region is presented in Table S1 of the supplementary material. The radius used for data retrieval was $r=15$ km, while the standard deviation of the Gaussian kernel was $\sigma=2.1$ km. The probability densities were thresholded at $f=0.3$ and the resulting candidate regions were pruned by setting $M_w=100$ m. We note that, for comparison purposes, we consider urban regions having a similar population, roughly in the range $[1\times 10^5,6\times 10^5]$. 



\section{Geometry characterization}

In this section we characterize the geometry of the urban regions by calculating their lacunarity~\cite{gefen1983geometric,allain1991characterizing}. Usually attributed to Mandelbrot~\cite{mandelbrot1983fractal}, the lacunarity has traditionally been used to characterize fractal structures~\cite{smith1996fractal,allain1991characterizing}. Nevertheless, the measurement has far reaching applications in different fields~\cite{plotnick1996lacunarity,tolle2003lacunarity,barros2008accuracy}. The lacunarity is commonly used to measure the spatial regularity of holes in objects represented in a binary image. For example, one can define a matrix $I$ containing the representation of a regular grid as an image, that is, $I_{ij}=1$ if a line of the grid passes through the point $(i,j)$, and $I_{ij}=0$ otherwise. This image can then be used as input to the lacunarity calculation. Since a regular grid can be regarded as a structure containing holes with exactly the same size, its lacunarity is close to the minimum value of the measurement, which is defined in the range $[1,\infty]$. A finite grid never reaches the minimum value due to border effects on the lacunarity calculation. This influence is lessened by applying a self-referred version of the lacunarity~\cite{rodrigues2004self}, which we use in the following.  

\begin{figure}[!h]
  \begin{center}
 \includegraphics[width=\columnwidth]{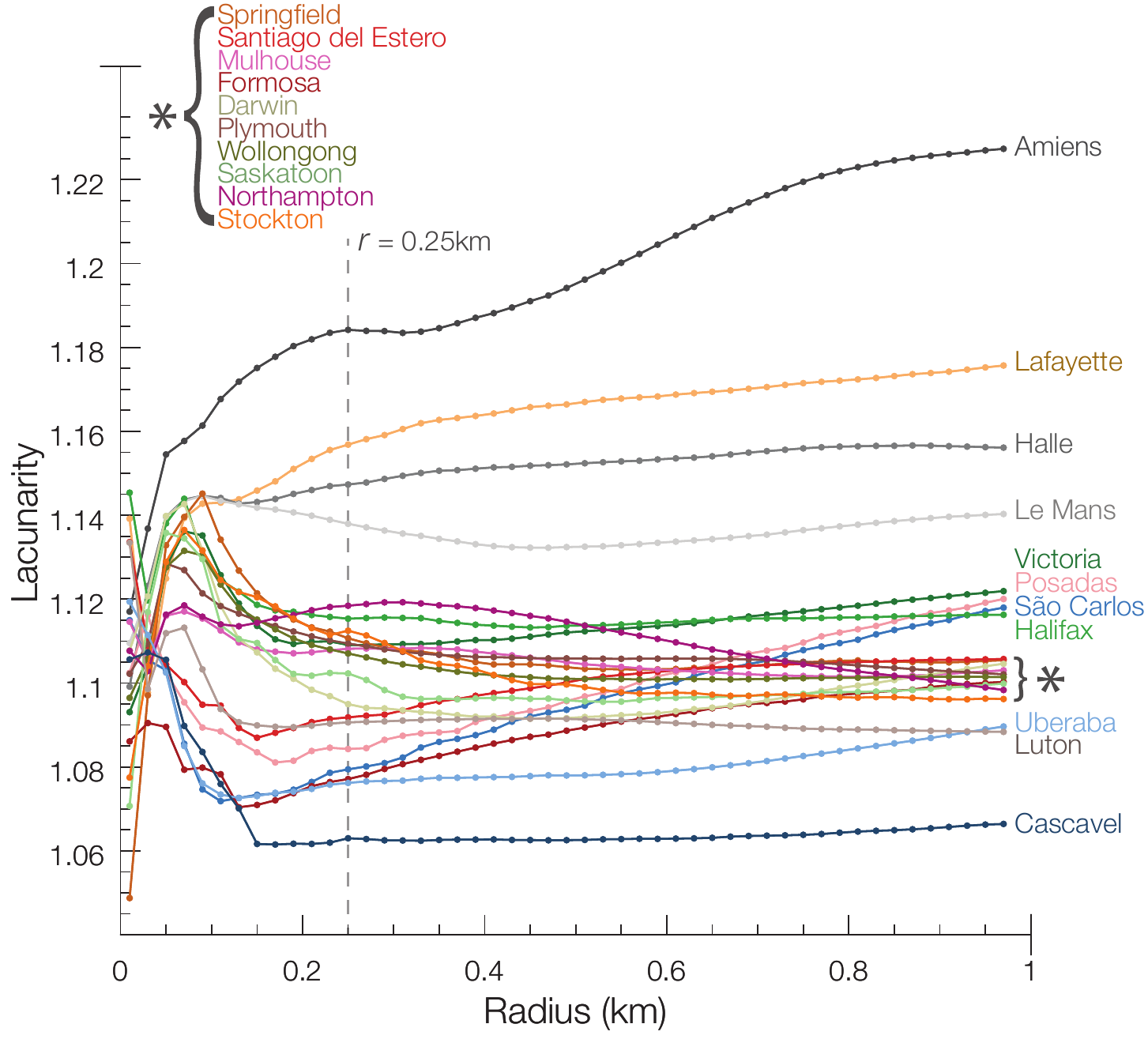} 
  \caption{Lacunarity of urban regions for different considered scales. The vertical dotted line represents the radius at which we rank the regions according to their lacunarity, which are then displayed in Figure \ref{f:city_visu_lacu}.}
  ~\label{f:lacu}
  \end{center}
\end{figure}

\begin{figure*}[!hp]
  \begin{center}
 \includegraphics[width=0.9\textwidth]{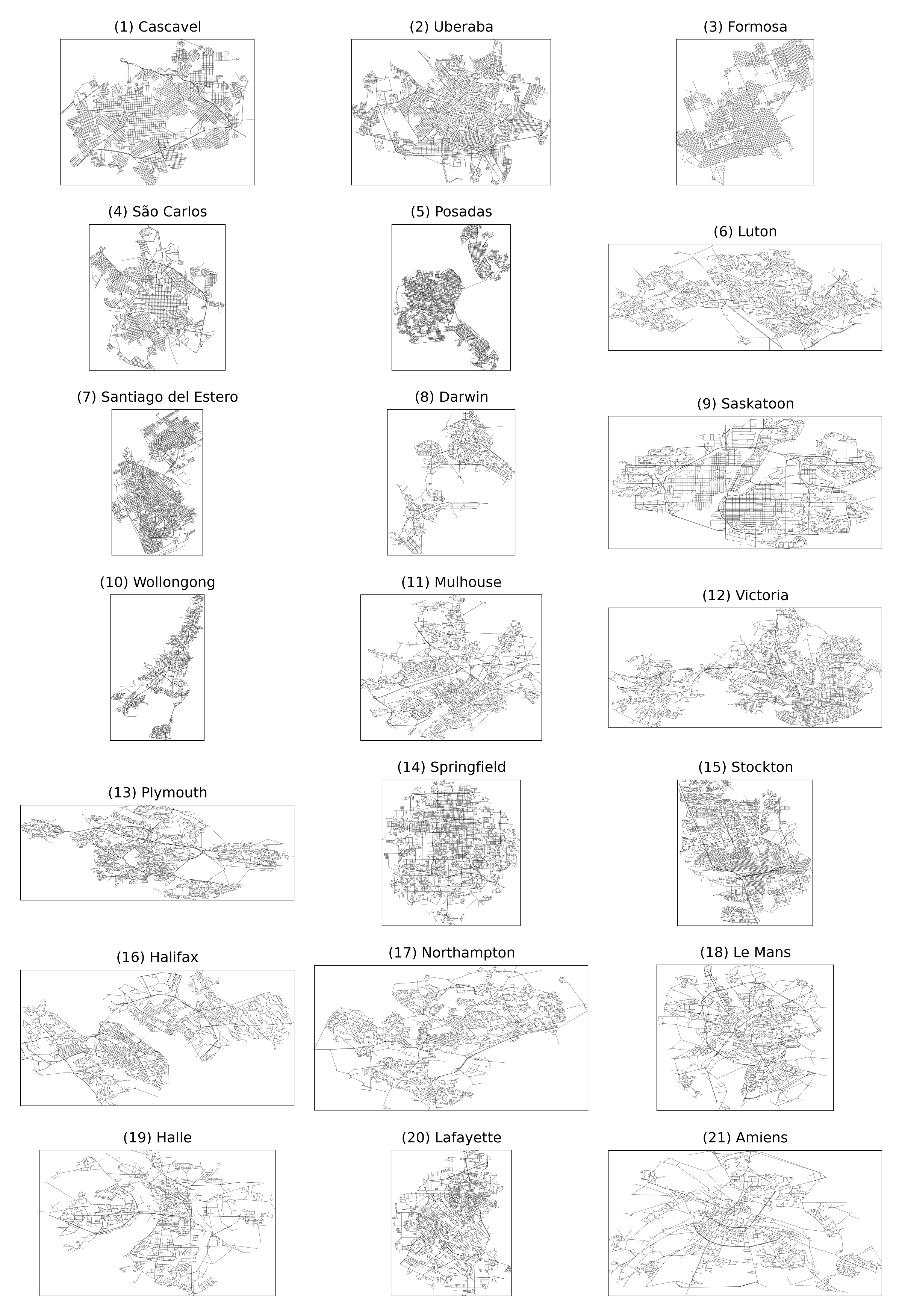} 
  \caption{Visualization of the urban regions structures, which are ordered according to increasing values of lacunarity.}
  ~\label{f:city_visu_lacu}
  \end{center}
\end{figure*}

The lacunarity is a multiscale property, where the scale is set by the radius $r$ of the neighborhood used to calculate hole size variations around each image pixel. In the case of urban regions, this means that we can look for regularity differences from the city block up to the entire city scale. For each urban region, we define its respective binary image as a drawing of the streets belonging to the region. The lacunarity is then calculated over the set of obtained images. In Figure \ref{f:lacu} we show the lacunarity of the urban regions as a function of the radius used in the calculation. It is clear that at very small scales ($r<100$ meters), the urban regions have similar lacunarity, which indicates similar regularity. This happens because, at this scale, we are only considering the structure of the roads. Around $r\approx 200$ meters there is already a defined regularity ranking between the regions. This is a scale that spans a few city blocks. For a very large scale ($r\approx 1000$ meters), the obtained ranking is mostly unchanged, with only a few urban regions having increased lacunarity. This happens because the contribution of the regions border to the lacunarity becomes significant, which decreases the estimated regularity of the structure.

Since the scale $r\approx 250$ meters seems to represent a good balance between statistical significance and absence of border effects, we consider the regularity hierarchy obtained at this scale. In Figure \ref{f:lacu} we indicate by a vertical dotted line the value $r=250$ meters. In Figure \ref{f:city_visu_lacu} we show each urban region ordered according to increasing values of lacunarity. The obtained ordering seems to reflect the notion that some cities are structured in a grid-like fashion~\cite{cardillo2006structural}, thus displaying low lacunarity values, while others have a seemingly disordered structure, having city blocks of many different sizes. 

\section{Topology characterization}

Another aspect that can be studied about the urban regions obtained with the presented methodology is the topological properties of their street networks. The OpenStreetMaps dataset contains information about the streets hierarchy, in the sense of expected traffic flow or projected traffic speed. Therefore, it is interesting to analyze how the edge betweenness~\cite{brandes2008variants} values of the streets are associated with their respective hierarchy. For such a task, we calculate the edge betweenness for three main classes of streets.

\begin{figure*}[]
  \begin{center}
 \includegraphics[width=\textwidth]{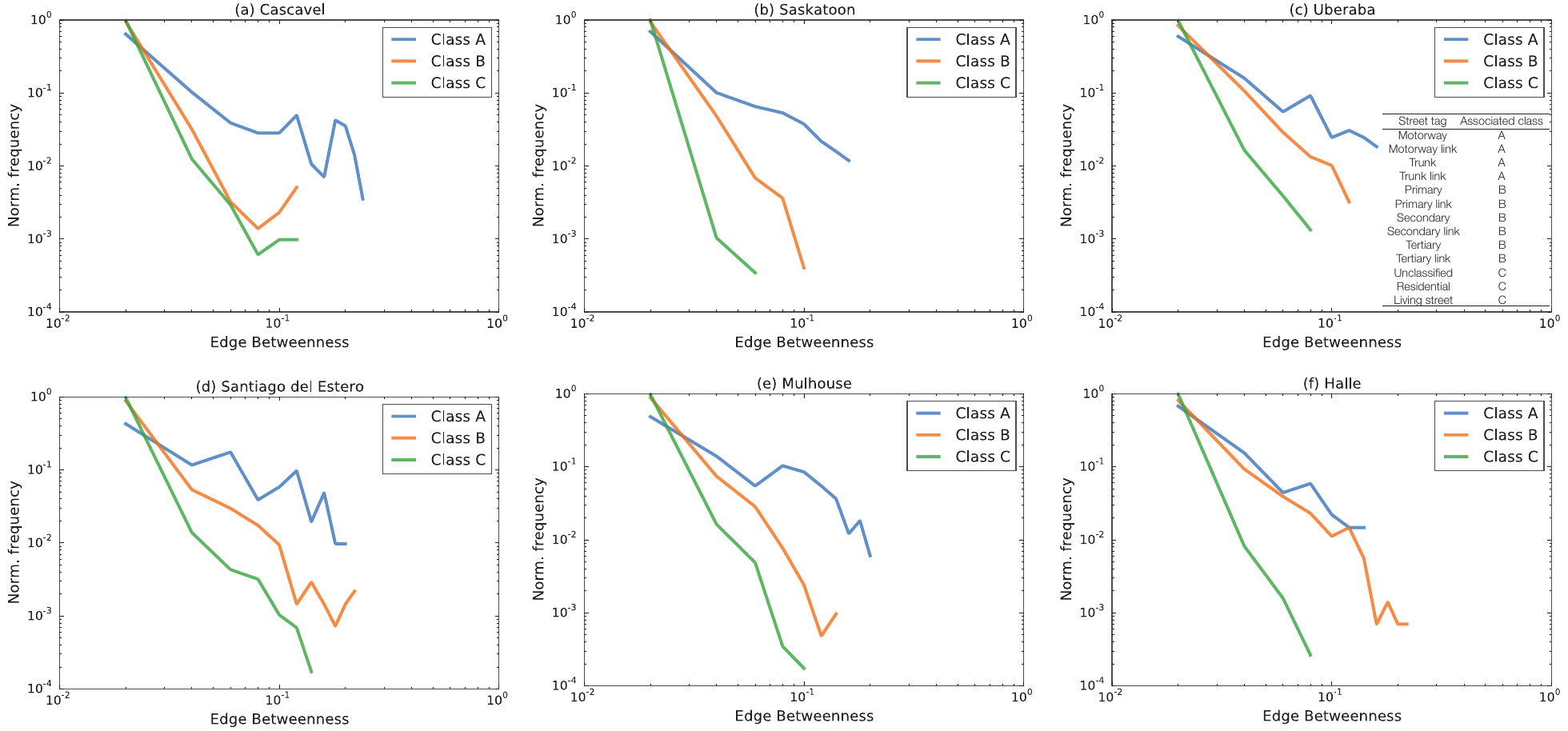} 
  \caption{Comparison of the edge betweenness and planned hierarchy of streets. The plots show the edge betweenness histograms calculated for six urban regions of the dataset. The values are separated into three classes, and their frequency is normalized individually for each class. The OpenStreetMaps tags and their respective associated class are indicated in panel (c).}
  ~\label{f:city_bet}
  \end{center}
\end{figure*}

We intuitively expect that the ``main'' streets, that is, streets planned to sustain larger traffic flows and speeds, will present larger betweenness values. In Figure~\ref{f:city_bet} we show the edge betwenness distribution for six urban regions. The OpenStreetMaps tags associated to the streets, as well as the respective three-class division considered are shown in Figure \ref{f:city_bet}(c). The histograms were normalized individually for each street class, since the number of streets in classes B and C is much larger than in class A. We note that the results shown in the figure are consistent for the other urban regions obtained using the methodology. It is clear that main streets, which are associated to class A, do indeed tend to have larger edge betweenness. 

The results also indicate that streets having low edge betweenness values do not necessarily belong to classes B or C. This is mostly caused by main streets having exits or intersections with B and C-class streets. The topological representation of intersections as nodes do not consider the preferential traffic flow associated with the main streets, and therefore the role of some main streets gets overlooked in the topological analysis. Nevertheless, it is interesting that many main streets do have large betwenness. This means that this measurement can be used to identify main streets when the metadata do not contain information about the streets hierarchy.

\section{Conclusion}

The very definition of what constitutes a city or urban region is strongly associated to its border demarcation. Therefore, it is expected that studies trying to derive universal laws for cities should give special attention to the actual area associated to the city. We believe that this stands true specially for studies in network theory, which commonly use administrative areas to delimit the networks.
 
We presented a simple approach to detect urban area borders based on an important structural aspect of cities, their road transportation network. The main advantage of the method is in its simplicity and intuitive behavior when changing each of its parameters. This is because each parameter involves a precise rule defining the requested border. The standard deviation of the Gaussian kernel, $\sigma$, sets the maximum typical distance between two sets of streets. The probability density fraction, $f$, used to threshold the density map of street intersections, establish the minimum density of street crossings. The parameter can be related to the admissible complexity of the streets configuration inside the urban area. The minimum width of the urban region, $M_w$, eliminates thin patches of streets that provide communication between two distinct urban regions.

Two possible applications allowed by the proposed methodology were discussed. The lacunarity measurement provided an interesting ranking of the geometrical regularity of the considered cities. It remains as an interesting study to identify the difference in lacunarity between planned cities and unplanned ones. Also, it would be interesting to verify the influence of the regions' relief (e.g. mountains, lakes and rivers) or urban architecture (e.g. parks, open spaces and large buildings) on the lacunarity. The topological analysis showed that the edge betweenness does indeed correlates with street importance, even when one considers a simple network construction defined by street crossings and terminations. Therefore, the betweenness could be used, for example, as a planning tool for constructing new streets.

As mentioned before, the presented methodology can be applied to any urban region in the world, provided there is data for such. It remains as a challenge to apply it in a world-wide scale in order to, among other interesting studies, search for classes of lacunarity, measure the topological regularity and test the legitimacy of Zipf's and Gibrat's laws for the urban areas.\\

\section*{Acknowledgments}

CHC thanks FAPESP (Grant no. 11/22639-8) for financial support. FNS acknowledges CAPES. LdFC thanks CNPq (Grant no. 307333/2013-2), NAP-PRP-USP and FAPESP (Grant no. 11/50761-2) for support.

\bibliographystyle{unsrt}
\bibliography{paper}

\newpage

\onecolumngrid

\section*{Supplementary information for ``A diffusion-based approach to obtaining the borders of urban areas''}

\renewcommand\thetable{S\arabic{table}} 
\setcounter{table}{0} 

\begin{table*}[h]
\caption{Urban areas considered in the main text for geometrical and topological characterizations. The name of the main city or region is indicated, together with the respective country.}
\begin{center}
\begin{tabular}{cc}
\hline
Urban area & Country \\
\hline
Stockton, California & United States \\
Lafayette, Louisiana & United States \\
Springfield, Missouri & United States \\
Uberaba, Minas Gerais & Brazil \\
Cascavel, Paran\'a & Brazil \\
S\~ao Carlos, S\~ao Paulo & Brazil \\
Halifax, Nova Scotia & Canada \\
Saskatoon, Saskatchewan & Canada \\
Victoria, British Columbia & Canada \\
Santiago del Estero, Santiago del Estero & Argentina \\
Posadas, Misiones & Argentina \\
Formosa, Formosa & Argentina \\
Plymouth, Devon & England \\
Luton, Bedfordshire & England \\
Northampton, Northamptonshire & England \\
Mulhouse, Alsace & France \\
Le Mans, Pays de la Loire & France \\
Amiens, Picardy & France \\
Halle, Saxony-Anhalt & Germany \\
Darwin, Northern Territory & Australia \\
Wollongong, New South Wales & Australia \\
\hline
\end{tabular}
\end{center}
\end{table*}

\end{document}